\documentclass[%
reprint,
 amsmath,amssymb,
 aps,
]{revtex4-1}

\usepackage{graphicx}

\usepackage[utf8]{inputenc}
\usepackage{amsmath,amssymb}

\usepackage{algorithm}
\usepackage[noend]{algpseudocode}

\DeclareMathOperator{\E}{\mathbb{E}}

\usepackage [english]{babel}
\usepackage [autostyle, english = american]{csquotes}

\begin{document}

\preprint{APS/123-QED}

\title{Coarse-Graining Auto-Encoders for Molecular Dynamics}

\date{March 2019}

\author{Wujie Wang}
\affiliation{%
 Department of Materials Science and Engineering\\
 Massachusets Institute of Technology\\
  77 Massachusetts Avenue, Cambridge, MA 02319
 }%
\author{Rafael G\'omez-Bombarelli}%
\affiliation{%
 Department of Materials Science and Engineering\\
 Massachusets Institute of Technology\\
 77 Massachusetts Avenue, Cambridge, MA 02319
}%
\email{rafagb@mit.edu}%

\begin{abstract}
Molecular dynamics simulations provide theoretical insight into the microscopic behavior of materials in condensed phase and, as a predictive tool, enable  computational design of new compounds. However, because of the large temporal and spatial scales involved in thermodynamic and kinetic phenomena in materials, atomistic simulations are often computationally unfeasible. Coarse-graining methods allow simulating larger systems, by reducing the dimensionality of the simulation, and propagating longer timesteps, by averaging out fast motions. Coarse-graining involves two coupled learning problems; defining the mapping from an all-atom to a reduced representation, and the parametrization of a Hamiltonian over coarse-grained coordinates. Multiple statistical mechanics approaches have addressed the latter, but the former is generally a hand-tuned process based on chemical intuition. Here we present Autograin, an optimization framework based on auto-encoders to learn both tasks simultaneously. Autograin is trained to learn the optimal mapping between all-atom and reduced representation, using the reconstruction loss to facilitate the learning of coarse-grained variables. In addition, a force-matching method is applied to variationally determine the coarse-grained potential energy function. This procedure is tested on a number of model systems including single-molecule and bulk-phase periodic simulations.

\end{abstract}

\maketitle

\section{Introduction}

Coarse-Grained (CG) molecular modeling has been used extensively to simulate complex molecular processes at a lower computational cost than all-atom simulations \cite{Agostino2017, Huang2010}. By compressing the full atomistic model into a reduced number of pseudo atoms, CG methods focus on the slow collective atomic motions and average out fast local motions. Current approaches generally focus on parametrizing coarse-grained potentials from atomistic simulations \cite{Noid2008} (bottom-up) or experimental statistics (top-down) \cite{Marrink2007, Periole2009}. The choice of all-atom to CG mapping plays an important role in recovering consistent CG dynamics, structural correlation and thermodynamics, \cite{Rudzinski2014, Noid2013} and a poor choice can lead to information loss in the description of slow collective interactions that are important for glass formation and transport. Systematic approaches to creating low resolution protein models based on essential dynamics have been proposed \cite{Zhang2008}, but a systematic bottom-up approach is missing for organic molecules of various sizes, resolutions and functionality. In general, the criteria for selecting CG mappings are usually based on \textit{a priori} considerations and chemical intuition.

Recently, machine learning tools have facilitated the development CG force fields \cite{Zhang_CG_2018, Bejagam2018, Lemke2017} and graph-based CG representations\cite{Webb2018, Chakraborty2018}. Here we propose to use machine learning to simultaneously optimize CG representations and potentials from atomistic simulations. One of the central themes in learning theory is finding optimal hidden representations from complex data sets \cite{Lawrence2005}. Such hidden representations can be used to capture the highest possible fidelity over complex statistical distributions with the fewest variables. We propose that finding the coarse-grained variables can be formulated as a problem of learning the latent variables in the atomistic data distribution. Recent works in unsupervised learning have shown great potential to uncover the hidden structure of complex data \cite{Tolstikhin2017,Kingma2014,Goodfellow2014}. As a powerful unsupervised learning technique, variational auto-encoders (VAEs) compress data through an information bottleneck \cite{Tishby2015} that continuously maps an otherwise complex data set into a low dimensional and easy-to-sample space. VAEs have been applied successfully to a variety of tasks, from image de-noising \cite{Vincent2010} to learning compressed representations for text \cite{Bowman2016}, celebrity faces, \cite{Liu2015} arbitrary grammars \cite{Kusner2017} or molecular structures \cite{Gomez-Bombarelli2018,Jin2018}. Recent works have applied VAE-like structures to learn collective molecular motions by reconstructing time-lagged configurations \cite{Wehmeyer2018} and hidden Markov models \cite{Mardt2018}.

\begin{figure*} 
    \centering
    \includegraphics[width=0.6\linewidth]{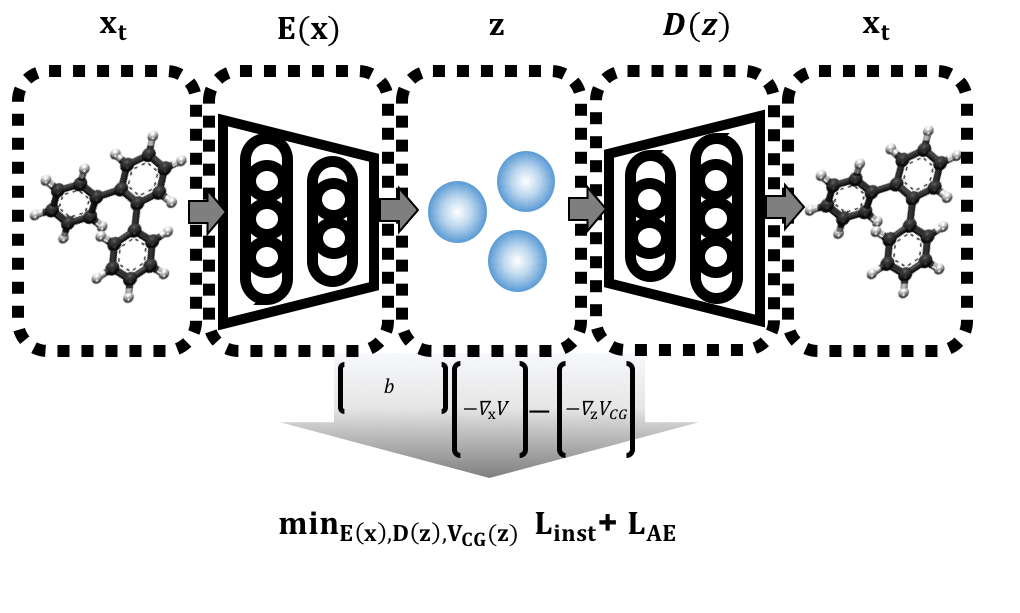}
    \caption{\textsf{Variational coarse-grain encoder framework. The model trains an auto-encoder that reconstructs the original all-atom data by encoding atomistic trajectories through a low dimensional bottleneck. A force-matching task can be simultaneously trained to find the CG mapping and force fields.}}
    \label{fig:diagram}
\end{figure*}

Here we apply an auto-encoder architecture (Figure \ref{fig:diagram}) with constraints to: 1) compress atomistic molecular dynamics (MD) data into a rigorously coarse-grained representation in 3D space; 2) train a reconstruction loss to help capture salient collective features from the all-atom data; and 3) adopt a supervised instantaneous force-matching approach to variationally find the optimal coarse-grained potential that matches the instantaneous mean force acting on the all-atom training data.

\section{Results}
Autograin is first demonstrated on coarse-graining single-molecule trajectories of ortho-terphenyl (OTP) and aniline ($\mathrm{C_6 H_7 N}$) in a vacuum. We initially train an auto-encoder for reconstruction and subsequently include the supervised force-matching task.


\begin{figure*}[tb]
    \centering
    \includegraphics[width=\textwidth]{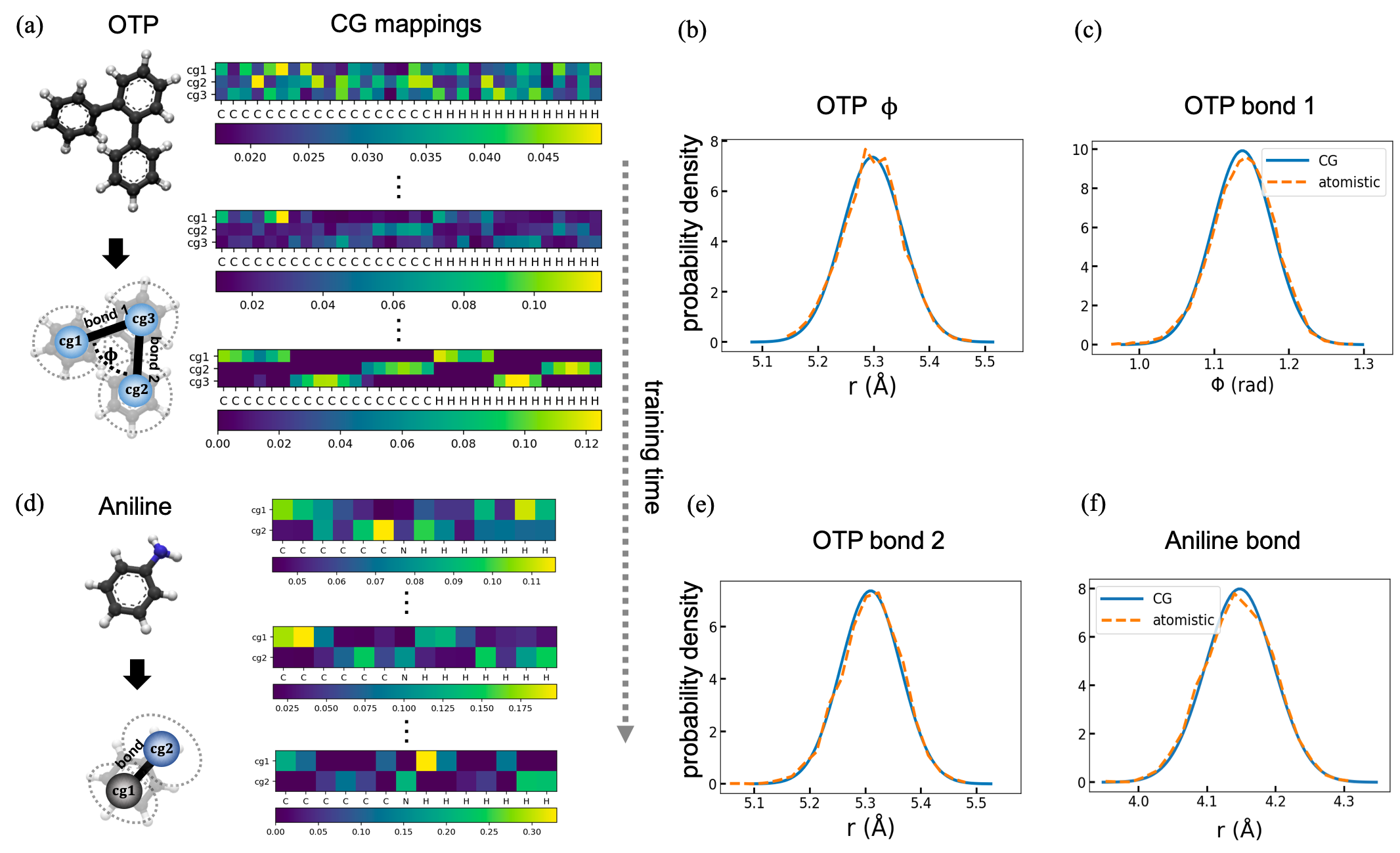}
    \caption{\textsf{\footnotesize{Learning of a CG mapping for OTP (a) and aniline (d) for increasing training time. The color bars represent the weights of the encoding mapping from all-atom to CG coordinates. Encoder weights are initialized with random weights and during the training process, the encoder is optimized to automatically coarse-grain atoms into pseudo-atoms. For the coarse-graining of OTP into three pseudo atoms, it automatically makes the coarse-graining decision of grouping each of the phenyl rings into three pseudo-atoms without human intervention. For coarse-graining aniline into two atoms, the coarse-graining decision learned is to group the $\mathrm{NH_2}$ group along with the two carbons and group the rest of the molecules into another pseudo-atom. As a validation for recovery of structural correlations, (b)(c)(f) show the corresponding bond and angle distribution of CG and mapped atomistic trajectories ofr OTP and (e) shows the bond distribution comparison between atomistic trajectories and CG trajectories for an aniline molecule. We plot the CG structural distribution by computing the normalized Boltzmann probability for the bond distribution: $P_{bond}(r) \propto \exp(\beta k_{bond} (r -r_0)^2)$, where $k_{bond}$ and $r_0$ are obtained from training the force-matching task.}}}
    \label{fig:singlemolecule_trainig}
\end{figure*}

For the OTP molecule, we choose $N = 3$ as the dimension of the coarse-grained space and each coarse-grained super-atom is treated as different species. The model is first initialized with random weights and trained as described in Algorithm \ref{alg:VCGE} by gradually increase the value $\beta$. The coarse-grain encoding gradually learned the most representative coarse-grained mapping by minimizing $L_{AE}$. For the case of OTP, the coarse-grained rules automatically captured by the model is to group each of the phenyl rings into one bead (Figure \ref{fig:singlemolecule_trainig} a). For the coarse-graining of aniline into two pseudo atoms, our model selects the coarse-grain mapping that partitions the molecules into two beads: one contains the amino group and the closest three phenyl carbons plus their attached hydrogens, the other groups three carbons and their associated hydrogens (Figure \ref{fig:singlemolecule_trainig} d). This optimal mapping is not necessarily the first intuitive mapping one could propose, a more immediate choice being one particle on the phenyl and one on the amino group, presumably the spherical shape imposed by the choice of force field favors similarly-sized beads.


 We then performed new calculations in the coarse-grained variables $z$ using $V_{CG}$ to obtain validation trajectories in CG space, and compared the equilibrium structural correlations with held-out data from the all-atom simulations. As shown in Figure \ref{fig:singlemolecule_trainig} (b), (c), and (e), the mapped atomistic distributions derived from $V$ agree well with the Boltzmann distribution derived from $V_{CG}$ for each degree of freedom in the case of OTP. Figure \ref{fig:singlemolecule_trainig} (f) shows good agreement between bond distributions for aniline.
 
 Generally in coarse-graining, an arbitrary highly-complex potential can be trained to reproduce radial distribution functions, often at the expense of non-physicality (multiple local minima in two-body potentials, repulsive regions in between attractive regions, etc). Our approach was able to learn simple harmonic potentials that should result in higher transferability. When a highly expressive neural potential is trained, the curves are reproduced almost exactly.


\begin{figure}[tb]
    \centering
    \includegraphics[width=3in]{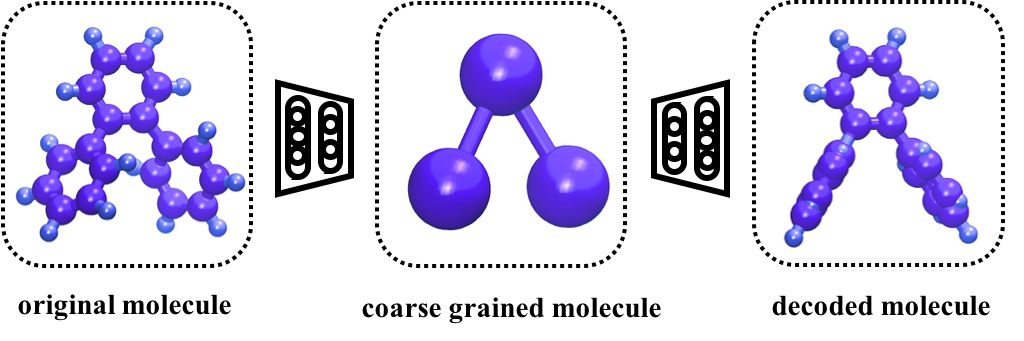}
    \caption{\textsf{\footnotesize{The demonstration of the encoded molecules and decoded molecules. The snapshot of original, encoded and decoded molecules are selected from real trajectories. The average reconstruction loss $L_{AE}$ is 0.406 ${\AA}^2$ . The two side phenyl rings in OTP are not reconstructed with full fidelity because the coarse-grained structures filter out detailed structural information. The decoded structures thus represent an averaged decoding.}}}
    \label{fig:reconstruction}
\end{figure}

Figure \ref{fig:reconstruction} shows a demonstration of the decoding of the OTP molecule. Because the coarse-graining encoding condenses the atomistic trajectories through an informational bottleneck, CG structures do not contain all the structural information in its original detail. By inspecting the decoded structure of OTP, we note that while the central phenyl rings can be decoded back with good fidelity, the two side phenyl rings however cannot be decoded back with original resolution. This is unsurprising, because the coarse-grained representation lacks the degrees of freedom to describe the relative orientations among phenyl rings. The coarse-grained super-atoms condense different relative rotation of the two side phenyl rings into the same coarse-grained states, and the information about rotational degrees of freedom is lost. Therefore, the decoder learns to map the coarse-grained variables into a averaged mean structure that represents the ensemble of relative rotations of the two side phenyl rings. The prospect of stochastic decoding functions to capture thermodynamic up-scaling is discussed below.

We have also applied Autograin on liquid system of methane and ethane. The training trajectories are for 64 all-atom molecules. The encoder and force-matching functional we trained as described as above. After training, the learned coarse-grained mapping and $V_{CG}$ was applied to coarse grain a test system of 512 methane and 343 ethane molecules with the same density. The relevant pairwise structural correlation functions for each individual system were then compared.

\begin{figure}[tb]
    \includegraphics[width=3in]{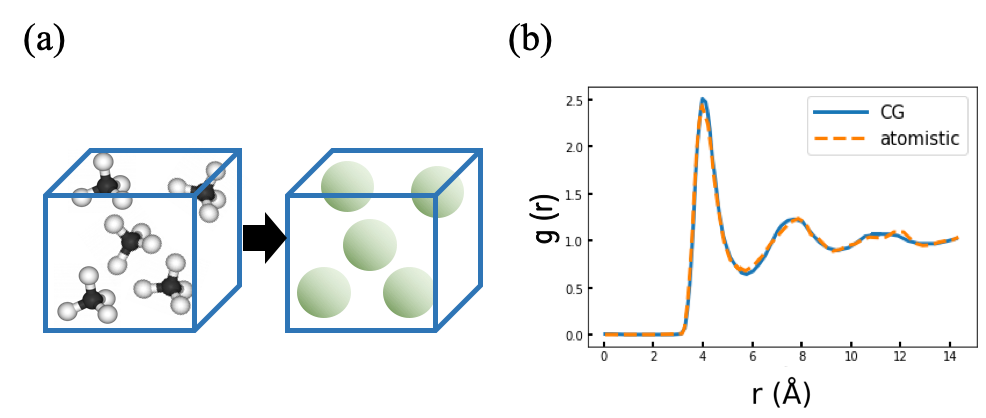}
    \caption{\textsf{\footnotesize{Comparison between CG and mapped atomistic pair correlation for methane liquids. Each methane molecule is coarse-grained into one pseudo-atom.}}}
    \label{fig:methane_liquid}
\end{figure}

\begin{figure}[tb]
    \includegraphics[width=3in]{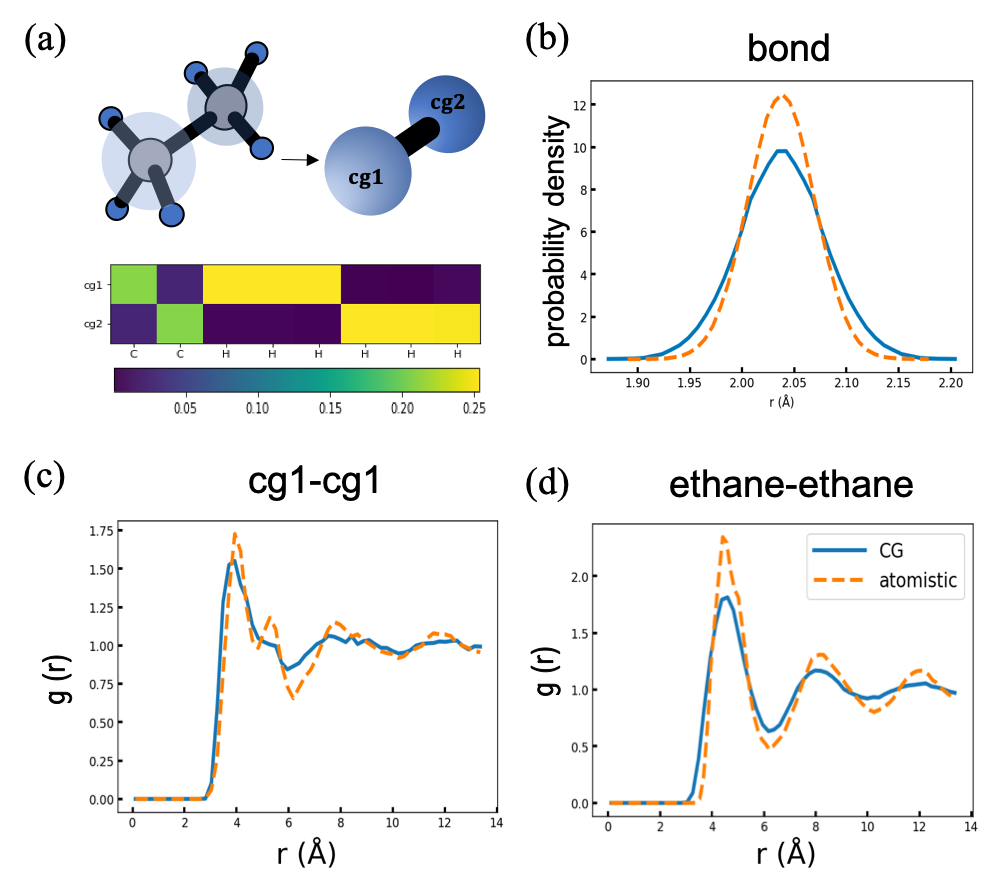}
    \caption{\textsf{\footnotesize{Pair correlation function validation using the learned CG classical force field. In this task, each ethane molecule is coarse-grained into two pseudo atoms. In $V_{CG}$, we choose the two pseudo atoms to be same type. a) Learned CG mapping. b-d) Comparison of CG and mapped structural correlation function.}}}
    \label{fig:ethane_liquid}
\end{figure}

\begin{figure}[tb]
    \includegraphics[width=3in]{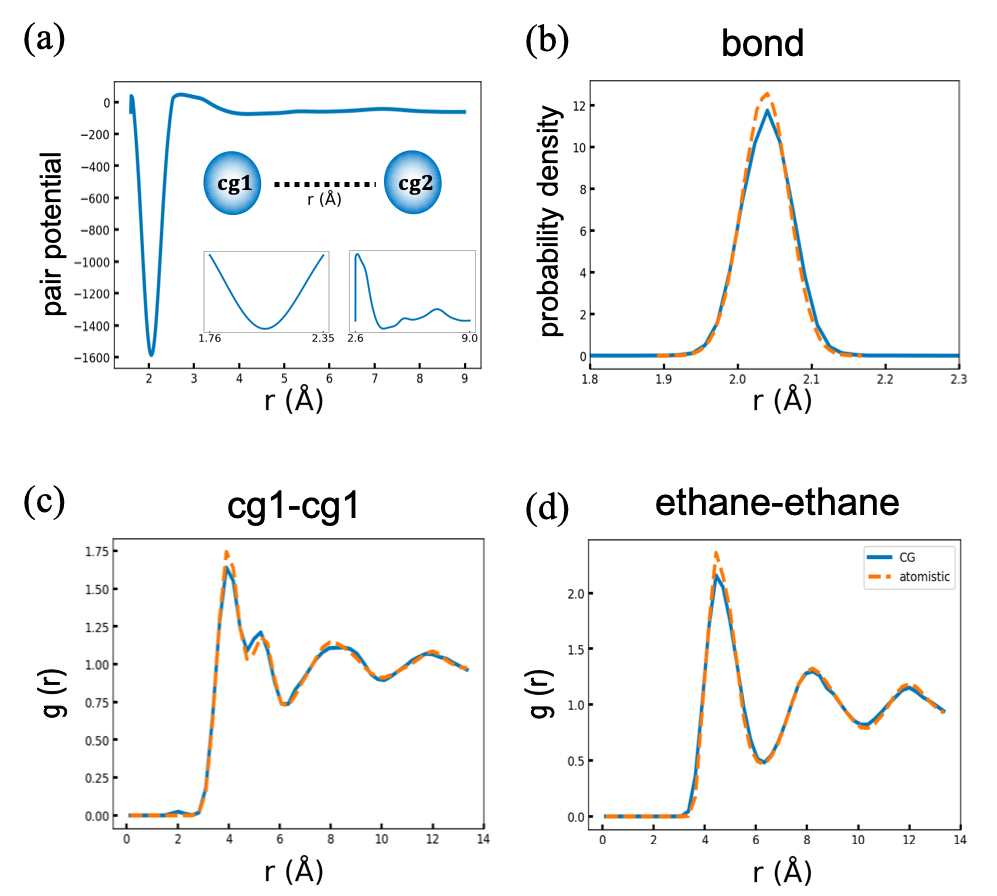}
    \caption{\textsf{\footnotesize{Pair correlation function validation using the learned CG neural force field. a) pairwise potential between two cg atoms extrapolated from the learned force field. The insets show that the potential approximates a harmonic bond potential at short distances; at longer distance, the potential approximates the non-bonded pair potential between cg atoms. b-d) The structural correlation functions generated from coarse grained neural potential show near-perfect agreement with the atomistic data.}}}
    \label{fig:ethane_neural}
\end{figure}

For methane, we choose $N = 1$ and only include 12-6 Lennard Jones interactions in $V_{CG}$. As shown in Figure\ref{fig:methane_liquid}, the correlation function of coarse-grained particles and obtain nearly perfect agreement between the CG and atomistic simulations. This is a expected result because the pairwise term is the only potential energy form in $V_{CG}$ and therefore there are no cross correlations between different energy terms.

For ethane, we choose $N =2$ and include the bonded potential and a 9-6 Lennard Jones potentials to describe the Van der Waals interactions in $V_{CG}$. From training, we obtain a coarse-grained mapping that groups each $\mathrm{CH_3}$ moiety into one pseudo atom. As seen in Figure \ref{fig:ethane_liquid}, reasonable agreement is obtained in the correlation function between the CG and mapped atomistic trajectories. We postulate that the discrepancy arises from a combination of: 1) The form of $V_{CG}$ only including classical bonded and non-bonded term, and thus lacking sufficient flexibility to fit any arbitrary interacting potentials. As discussed above, during coarse-graining it is common to compensate the high-order correlations lost from spatial coordinates into complex contributions to the potential that represent higher-order effects. 2) The force-matching method does not address structural cross correlation and it is not necessarily guaranteed to recover the atomistic correlation function perfectly, as discussed by Noid \textit{et al.} and Lu \textit{et al.} \cite{Noid2013, Lu2013}. The structural cross-correlation consideration is addressed in other CG methods like generalized Yvon-Born-Green method \cite{Mullinax2010} and iterative force matching \cite{Lu2013}.

In order to explore the role of those two sources of error, we trained a neural potential expression of $V_{CG}$ This more expressive potential makes up the shortcomings of the simple functional form. The model we implemented is based on Deep Tensor Neural Network \cite{Schutt2018, Schutt2017}, an architecture that demonstrates the state-of-the-art accuracy to fit potential energy surfaces from quantum chemistry calculations. The model constructs highly complex correlations among coarse-grained beads by iteratively convolving over the 3D point clouds to produce an energy given input coarse-grained coordinates. Because such neural network models are fully differentiable functions, forces can be readily evaluated by taking the derivatives so that we can directly apply force matching. Due to its inherent ability to capture hight order correlations, the neural force field can learn the many-body potential of mean force that almost exactly reproduces the mapped structural correlation function as shown in Figure \ref{fig:ethane_neural} where the neural potential learns a complicated function to (Figure \ref{fig:ethane_neural}(a) insets) to better reproduce the pair correlation function and the model inherently incorporates the cross-correlation between different degrees of freedom.

\section{Discussion}

Within the current framework, there are several possibilities for future research directions, regarding both the supervised and unsupervised parts. 

Here, we have presented a choice of deterministic encoder and decoder (see method section). However, such a deterministic CG mapping results, by construction, in an irreversible loss of information. This is reflected in the reconstruction of average all-atom structures instead of the reference instantaneous configurations. By combining with predictive inference for atomistic back-mapping \cite{Schoberl2017}, a probabilistic auto-encoder can go further by learning a reconstruction probability distribution that reflects the thermodynamics of the degrees of freedom averaged out by the coarse-graining. Using this framework as a bridge between different scales of simulation, generative models can help build better hierarchical understanding of multi-scale simulations.

Whereas the choice of force-matching approach does not guarantee the recovery of individual pair correlation functions derived from full atomistic trajectories \cite{Lu2013, Noid2013}, but we have observed that with simple functional forms most of the error arises from the lack of complex terms in the potential of mean force, and the use of neural potentials recovers  . To include the learning of structural cross-correlations, our method can optimized to incorporate iterative force matching \cite{Lu2013} and relative entropy \cite{Shell2016}.  

The automatic learning of multi-particle force fields on the fly requires automatic classification of atoms and variationally building empirical force-field topologies at training time.  In the current model, a pre-determined topology is needed to calculate the total potential energy. It would be ideal to develop a probabilistic way to generate force field topologies for discrete particle types that are variationally optimized along coarse-graining encoding. Recent advances in learning graphs shed some light in this line of research \cite{Jin2018,Xie2018,duvenaud_convolutional_2015}.

Methods based on force-matching, like other bottom-up approaches such as relative entropy, attempt to reproduce structural correlation functions at one point in the thermodynamic space. As such, they are not guaranteed to capture non-equilibrium transport properties \cite{Noid2013, Davtyan2015} and are not necessarily transferable among different thermodynamic conditions \cite{Noid2013, Carbone2008, Krishna2009}. The data-driven approach we propose enables learning over different thermodynamic conditions. In addition, this framework opens new routes to understanding the coarse-grained representation influences transport properties by training on time-series data. A related example in the literature is to use to use a time-lagged auto-encoder \cite{Wehmeyer2018} to learn a latent representation that best captures molecular kinetics.

 In summary, we propose to treat coarse-grained coordinates as latent variables which can be sampled with molecular dynamics. By regularizing the latent space with force matching, we jointly train the encoding mapping, a deterministic decoding, and a transferable potential that can be used to simulate larger systems for longer times and thus accelerate molecular dynamics. Our work also opens up possibilities to use statistical learning as a basis to bridge across multi-scale simulations.

\section{Methods}

Autograin is based on a semi-supervised learning approach based on auto-encoders to create an all-atom to CG mapping function, as well as a potential in CG coordinates that can later be used to carry out new simulations for larger systems with lower computational cost. The latent structure is shaped by training both an unsupervised reconstruction task and a supervised force-matching task. To learn corresponding force fields that can be transferred, the model carries out a variational coarse-grained force matching that incorporates the learning of the coarse-grained mapping in the force-matching functional.

\subsection{Coarse-Graining Auto-encoding}
Noid \textit{et al.} have studied the general requirements for a physically rigorous mapping function \cite{Noid2008}. In order to address those requirements, Autograin is trained to optimize the reconstruction of atomistic configurations by propagating them through a low-dimension bottleneck in Cartesian coordinates. Unlike most instances of VAEs, the dimensions of the CG latent space have physical meaning. Since the CG space needs to represent the system in position and momentum space, latent dimensions need to correspond to real-space Cartesian coordinates and maintain the structural information of molecules. 

We make our encoding function a linear mapping in Cartesian space $E(x): \mathbb{R}^{3n} \rightarrow \mathbb{R}^{3N} $ where n is the number of atoms and N is the desired number of coarse-grained particles.

Let $x$ be atomistic coordinates and z be the coarse-grained coordinates. The encoding function should satisfy the following requirements \cite{Darve,Noid2008}:

\begin{enumerate}
\item  $z_i = E (x) = \sum^n_{j =1} E_{ij} x_j \in \mathbb{R}^3 , i = 1 \ldots N,  j = 1 \ldots n $
\item  $\sum_j E_{ij}= 1 $ and $E_{ij} \geq 0$
\item  Each atom contributes to at most one coarse-grained variable $z$
\end{enumerate}

where $E_{ij}$ is the matrix element in , $j$ is the index for atoms, $i$ is the index for coarse-grained atoms. Requirement (2) defines the coarse-grained variables to be the statistical averages of the Cartesian coordinates of contributing atoms.  In order to maintain the consistency in the momentum space after the coarse-grained mapping, the coarse-grained masses are rigorously redefined as $M_{z} = (\nabla E(x)^{-1})^\intercal M  \nabla E(x)^{-1}$ \cite{Darve, Noid2008}. And this definition of mass is a corollary of requirement (3). 

To specifically satisfy requirement (3), we design the encoder based on Gumbel-Softmax \cite{Jang2016} with a tunable fictitious 
``temperature" that can be adjusted during the training to learn discrete variables. The detailed algorithm is described as in Algorithm 1.

The softmax function is thus used to ensure that the encoding function represents the atomic contributions for each of the coarse-grained pseudo atoms. We apply the Gumbel-Softmax function with a fictitious inverse ``temperature" $\beta$ on a separate weight matrix which is used as a mask on the encoding weight $E_{ij}$ matrix. By gradually increasing $\beta$ toward a sufficiently high inverse ``temperature",  the mask will asymptotically choose only one coarse-grained variable for each of the atom which satisfies requirement (3). This is equivalent to an attention mechanism, which is widely used in deep learning \cite{Vaswani2017}.

The decoding of coarse grained pseudo atoms has received little attention in the literature, so we opt for a simple decoding approach. Thus, we use a matrix $D$ of dimension $n $ by $ N $ that maps coarse-grained variables back to the original space. Hence, both the encoding and decoding mappings are deterministic. Although deterministic reconstruction via a low dimensional space leads to irreversible information loss, the decoder and encoder functions are sufficient to construct the information bottleneck to learn the latent representation of molecular conformations in coarse-grained coordinates. The unsupervised optimization task is to minimize the reconstruction loss:

\begin{equation} \label{eq:11}
\min_{D, E}L_{AE} = \min_{D, E} \E[(D E(x_t) - x_t)^2]
\end{equation}

\begin{algorithm}[H]
\caption{Variational Coarse-grained Encoding}\label{alg:VCGE}
\begin{algorithmic}

\State $A_{ij}, B_{ij}, D_{ji}, \beta,  \Delta \beta \gets$ initialize parameters
\Repeat{}

\State $X \gets$ random mini-batch molecular dynamics frames
\State $E_{ij}\gets \frac{e^{A_{ij}}}{\sum_j^n e^{A_{ij}}}$
\State $B_{ij}\gets \frac{e^{B_{ij} \beta}}{\sum_j^n e^{B_{ij} \beta}}$
\State $E_{ij}\gets E_{ij} \circ B_{ij}$
\State $E_{ij}\gets \frac{E_{ij}}{\sum_i^n E_{ij}}$

\State $ g \gets \nabla_{A_{ij}, B_{ij}, D_{ji}} L_{AE}(E_{ij}, D_{ji}; X) $
\State $A_{ij}, B_{ij}, D_{ji} \gets$ update parameters using gradients $g$ 
\State $\beta \gets \beta + \Delta \beta $
\Until convergence of $L_{AE}$
\end{algorithmic}
\end{algorithm}

\subsection{Variational Force Matching}
The CG auto-encoder provides an unsupervised variational method to learn the coarse grained coordinates. In order to learn the coarse-grained potential energy $V_{CG}$ as a function of also-learned coarse grained coordinates, we propose an instantaneous force-matching functional that is conditioned on the encoder. The proposed functional enables the learning of empirical force fields parameters and the encoder simultaneously by including the optimization of $E(x)$ in the force-matching procedure. Training empirical potentials from forces has a series of advantages: (i) the explicit contribution on every atom is available, rather than just pooled contributions to the energy, (ii) it is easier to learn smooth potential energy surfaces and energy-conserving potentials \cite{Chmiela2017} and (iii) instantaneous dynamics, which represent a trade-off in coarse-graining, can be captured better. Forces are always available if the training data comes from molecular dynamics simulations, and for common electronic structure methods based on density functional theory, forces can be calculated at nearly the same cost as self-consistent energies.

The force-matching approach builds on the idea that the average force generated by the coarse grained potential $V_{CG}$ should reproduce the coarse-grained atomistic forces from the thermodynamic ensemble \cite{Izvekov2005, Zhang_CG_2018, Ciccotti2005}. Given an atomistic potential energy function $V(x)$ with the partition function being $Z$, the probabilistic distribution of atomistic configurations is:
\begin{equation} \label{eq:1}
p(x) = \frac{1}{Z} e^{- \beta V(x)}
\end{equation}
The distribution function of coarse-grained variables $p(z)$ and corresponding many-body potential of mean force $A(z)$ are:
\begin{equation} \label{eq:2}
p(z) = \frac{1}{Z} \int e^{- \beta V(x)} \delta(E(x) - z) dx
\end{equation}
\begin{equation} \label{eq:3}
A(z) = -\frac{1}{\beta} \ln P(z)
\end{equation}

The mean force of the coarse-grained variables is the average of instantaneous force conditioned on $E(x) = z$ \cite{Kalligiannaki2015, Darve} assuming the coarse grained mapping is linear:

\begin{equation} \label{eq:4}
-\frac{dA}{dz} = F(z)  = \langle - b \cdot \nabla V(x) \rangle_{E(x) = z}
\end{equation}
\begin{equation} \label{eq:5}
 b = \frac{w}{w \cdot \nabla E(x)}
\end{equation}

where $F(z)$ is the mean force and $b$ represents a family of possible vectors such that $w \cdot \nabla E(x) \neq 0 $.  We further define $F_{inst}(z) = - b \cdot  \nabla V(x) $ to be the instantaneous force and its conditional expectation is equal to the mean force $F(z)$. It is important to note that $F_{inst} (z)$ is not unique and depends on the specific choice of $w$ \cite{Kalligiannaki2015, Ciccotti2005, DenOtter2000}, but their conditional averages return the same mean force.

For possible $b$, we further choose $w = \nabla E(x) ^\intercal$ which is a well-studied choice \cite{Ciccotti2005, DenOtter2000}, so that: 

\begin{equation} \label{eq:7}
b = \frac{\nabla E(x) ^ \intercal}{\nabla E(x) ^\intercal \cdot \nabla E(x)}
\end{equation}

With $b$ as a function of $\nabla E(x)$, we adopt the force-matching scheme introduced by Izvekov \textit{et al.} \cite{Izvekov2006, Izvekov2005}, in which the mean square error is used to match the mean force and the ``coarse-grained force" is the negative gradient of the coarse-grained potential. The optimizing functional, developed based on Izvekov \textit{et al.}, is 
\begin{equation} \label{eq:8}
\min_{\theta, E}L = \min_{\theta, E} \E[(F(z) + \nabla_z V_{CG}(E(x)))^2]
\end{equation}
where $\theta$ is the parameters in $V_{CG}$ and $\nabla V_{CG}$ represents the ``coarse grained forces" which can be obtained from automatic differentiation as implemented in open-source packages like PyTorch \cite{paszke2017automatic}. However, to compute the mean force $F$ would require constrained dynamics \cite{Ciccotti2005} to obtain the average of the fluctuating microscopic force. 
According to Zhang et al \cite{Zhang_CG_2018}, the force-matching functional can be alternatively formulated by treating the instantaneous mean force as an instantaneous observable with a well-defined average being the mean force $F(z)$:
\begin{equation} \label{eq:9}
F_{inst}(z) = F(z) + \epsilon(z)
\end{equation}
on the condition that $\E_z[F_{inst}] = F(z)$. 

Now the original variational functional becomes instantaneous in nature and can be reformulated as the following minimization target:
\begin{equation} \label{eq:10}
\min_{\theta, E} L_{inst} = \min_{\theta, E} \E[F_{inst}(z)+ \nabla V_{CG} (E(x)))^2]
\end{equation}

Instead of matching mean forces that need to be obtained from constrained dynamics, our model minimizes $L_{inst}$ with respect to $V_{CG}(z)$ and $E(x)$. $L_{inst}$ can be shown to be related to $L$ with some algebra : $L$: $L_{inst} = L + \E[\epsilon(E(x))^2]$ \cite{Zhang_CG_2018}. This functional provides a variational way to find a CG mapping and its associated force fields functions.

\subsection{Model Training}
The overall loss function to be optimized is the joint loss of the reconstruction loss and instantaneous force-matching loss. The total loss function is $L_{VCGE} = L_{AE} + L_{inst}$. The schematic for optimization stack is shown in Figure \ref{fig:diagram}.

We train the model from atomistic trajectories with the atomistic forces associated with each atom at each frame. The model is trained to minimize the reconstruction loss $L_{AE}$ along with force-matching loss $L_{inst}$ as shown in Figure \ref{fig:diagram}. It is propagated in the feed-forward direction and its parameters are optimized using back-propagation \cite{Hecht-Nielsen1989}. 

In practice, we first train the auto-encoder in an unsupervised way to obtain a representative coarse-graining mapping. The supervised force-matching task is then trained jointly with the auto-encoder to variationally find $V_{CG}$ and further optimize $E(x)$ and $D(z)$ to achieve a final coarse-grain mapping and its associated force fields.

We tested two general choices of functional forms. Simple potentials based on classical interactions like Lennard-Jones and harmonic bonded terms, as well as neural-network potentials. The former are fast to evaluate and transferable between different systems. They are a common choice when building simple phenomenological models or to carry out virtual screening since have the speed and scaling desired for large-scale CG simulations. However, they may not be sufficiently expressive to fit the complicated many-body potential of mean force. Neural network potentials \cite{Behler2007}, and traditionally spline potentials, have more flexibility in fitting and can capture the highly-complex potential of mean force more faithfully, but at the cost of poor transferability. 

\subsection{Computational details}
Molecular trajectory data was obtained by using the OPLS force field generated using the LigParGen server \cite{Dodda2017}. For gas-phase single-molecule trajectories, we use a 6-ps trajectory of 3000 frames obtained by Langevin dynamics with a friction coefficient of 1 $\mathrm{ps}^{-1}$. Models were pre-trained for 100 epochs with mini-batches of size 10 in the unsupervised. The Adam optimizer  \cite{Kingma2014} was used. We used PyTorch for training our model \cite{paszke2017automatic}.

For single-molecule OTP we learn a classical potential consisting of two bonds and angle, we train the force-matching task to find the harmonic bond and angle potential parameters that best matches the forces from the training data. The CG structural distribution is obtained by computing the normalized Boltzmann probability for the bonds and angle distributions: $p_{bond}(r) \propto \exp(\beta k_{bond} (r -r_0)^2)$ and $p_{\phi}(\phi) \propto \exp(\beta k_{\phi} (cos(\phi) -cos(\phi_0))^2)$ where $k_{bond}$, $k_{\phi}$, $r_0$ and $\phi_0$ are obtained from training the CG potential.

In the case of molecular liquids, the training trajectories are obtained using NVT at 100K for 64 methane molecules and 120K for 64 ethane molecules. The neural network potential shown in Figure \ref{fig:ethane_neural} is implemented based on SchNet, which is a variant of deep tensor neural net framework with continuous filters\cite{Schutt2018} and we apply two convolutions to fit the coarse-grained potential. The model is trained based on the CG mapping shown in Figure \ref{fig:ethane_liquid}(a). The validation run is done using Langevin Dynamics at 120K to simulate the NVT ensemble for coarse-grained ethane.

\section{Acknowledgements}
WW thanks Toyota Research Institute for financial support. RGB thanks MIT DMSE and Toyota Faculty Chair for support. WW and RGB thank Prof. Adam P. Willard (Massachusetts Institute of Technology) for helpful discussions.

\bibliography{main.bib}

\end{document}